\documentclass[sigconf]{acmart}

\usepackage[utf8]{inputenc} 
\usepackage[T1]{fontenc}    
\usepackage{hyperref}       
\usepackage{url}            
\usepackage{booktabs}       
\usepackage{amsfonts}       
\usepackage{nicefrac}       
\usepackage{microtype}      

\usepackage{algorithm}
\usepackage[noend]{algpseudocode}
\usepackage{algorithmicx}
\usepackage{subfigure}
\usepackage{soul}
\usepackage{graphicx}
\usepackage{amsmath,amsthm}
\usepackage{xcolor,soul}
\usepackage{tabularx}
\usepackage{verbatim}
\usepackage{balance,flushend}
\usepackage[normalem]{ulem}
\usepackage{balance}
\usepackage[multiple]{footmisc}
\usepackage{capt-of}
\usepackage[normalem]{ulem}

\providecommand{\keywords}[1]{\textbf{\textit{Keywords:}} #1}
\newcommand{\methodname}{CLRec }

\newcommand{\eat}[1]{}

\newcommand{\printfnsymbol}[1]{%
    \textsuperscript{\@fnsymbol{#1}}%
}

\newtheorem{theorem}{Theorem}
\theoremstyle{definition}

\newtheorem{remark}{Remark}

\AtBeginDocument{%
    \providecommand\BibTeX{{%
        \normalfont B\kern-0.5em{\scshape i\kern-0.25em b}\kern-0.8em\TeX}}}

\copyrightyear{2021}
\acmYear{2021}
\setcopyright{acmlicensed}\acmConference[KDD '21]{Proceedings of the 27th ACM SIGKDD Conference on Knowledge Discovery and Data Mining}{August 14--18, 2021}{Virtual Event, Singapore}
\acmBooktitle{Proceedings of the 27th ACM SIGKDD Conference on Knowledge Discovery and Data Mining (KDD '21), August 14--18, 2021, Virtual Event, Singapore}
\acmPrice{15.00}
\acmDOI{10.1145/3447548.3467102}
\acmISBN{978-1-4503-8332-5/21/08}
\begin{document}
    \fancyhead{}
    \title{
        Contrastive Learning for Debiased Candidate Generation in Large-Scale Recommender Systems
    }



    \author{Chang Zhou}
    \authornote{Equal contribution.}
    \affiliation{\institution{DAMO Academy, Alibaba Group}\city{}\country{}}
    \email{ericzhou.zc@alibaba-inc.com}

    \author{Jianxin Ma}
    \authornotemark[1]
    \authornote{Corresponding author.}
    \affiliation{\institution{DAMO Academy, Alibaba Group}\city{}\country{}}
    \email{jason.mjx@alibaba-inc.com}

    \author{Jianwei Zhang}
    \authornotemark[1]
    \affiliation{\institution{DAMO Academy, Alibaba Group}\city{}\country{}}
    \email{zhangjianwei.zjw@alibaba-inc.com}

    \author{Jingren Zhou}
    \affiliation{\institution{DAMO Academy, Alibaba Group}\city{}\country{}}
    \email{jingren.zhou@alibaba-inc.com}

    \author{Hongxia Yang}
    \affiliation{\institution{DAMO Academy, Alibaba Group}\city{}\country{}}
    \email{yang.yhx@alibaba-inc.com}

    \renewcommand{\shortauthors}{Zhou et al.}

    \begin{abstract}
        Deep candidate generation (DCG) that narrows down the collection of relevant items from billions to hundreds via representation learning has become prevalent in industrial recommender systems.
        Standard approaches approximate maximum likelihood estimation (MLE) through sampling for better scalability and address the problem of DCG in a way similar to language modeling.
        However, live recommender systems face severe exposure bias and have a vocabulary several orders of magnitude larger than that of natural language, implying that MLE will preserve and even exacerbate the exposure bias in the long run in order to faithfully fit the observed samples. 
        In this paper,
        we theoretically prove that a popular choice of contrastive loss is equivalent to reducing the exposure bias via inverse propensity weighting, which provides a new perspective for understanding the effectiveness of contrastive learning.
        Based on the theoretical discovery,
        we design \emph{CLRec}, a \emph{C}ontrastive \emph{L}earning method to improve DCG in terms of fairness, effectiveness and efficiency in \emph{Rec}ommender systems with extremely large candidate size.
        We further improve upon CLRec and propose Multi-CLRec, for accurate multi-intention aware bias reduction.
        Our methods have been successfully deployed in
        \emph{Taobao}, where at least four-month online A/B tests and offline analyses demonstrate its substantial improvements, including a dramatic reduction in the Matthew effect.
    \end{abstract}

\ccsdesc[500]{Information systems~Recommender systems}

    \keywords{Recommender systems; candidate generation; bias reduction; inverse propensity weighting; contrastive learning; negative sampling}

    \maketitle

    \section{Introduction}

Large-scale industrial recommender systems usually adopt a multi-stage pipeline, where the first stage, namely candidate generation, is responsible for retrieving a few hundred relevant entities from a billion scale corpus.
Deep candidate generation (DCG)~\citep{youtube}, a paradigm that learns vector representations of the entities to enable fast k-nearest neighbor retrieval~\citep{faiss}, has become an essential part of many live industrial systems with its enhanced expressiveness.

Typical large-scale DCG models~\citep{youtube,airbnb,tdm,mind} regard the problem of identifying the most relevant items to the users as estimating a multinomial distribution traversing over all items for each user, conditional on the user's past behaviors.
Maximum likelihood estimation (MLE) is the conventional principle for training such models.
Apparently, exact computation of the log likelihood, which requires computing softmax looping over a million- or even billion-scale collection of items, is computationally infeasible.
Among the various sampling-based approximation strategies, sampled-softmax~\citep{sampledsoftmax_raw,sampledsoftmax} usually outperforms the binary-cross-entropy based approximations such as NCE~\citep{nce} and negative sampling~\citep{w2v} at large scale.

However, the MLE paradigm and the sampling strategies mainly stem from the language modeling community, where the primary goal is to faithfully fit the observed texts.
Indeed, live recommender systems are different from natural language texts in several aspects.
Firstly, the training data are collected from current undergoing systems that might be sub-optimal and biased towards popular items.
In such situations, some high-quality items can be under-explored in the training data, while an algorithm trained via MLE will continue to under-estimate the relevance of the under-explored items in order to faithfully fit the observed data.
Secondly, the set of items for recommendation is much larger than the vocabulary of natural languages, e.g.\ $\approx 100M$ items in our system compared to $\approx 100k$ words, posing great difficulties in learning a set of fairly good representations for the entire candidate set.

In this paper, we introduce \emph{CLRec}, a \emph{C}ontrastive \emph{L}earning framework for debiased DCG in \emph{REC}ommender systems.
Contrastive learning, which constructs self-supervised tasks to improve the discriminative ability of a model, has become increasingly popular recently in the pre-training community~\citep{moco,simclr}.
We discover that the contrastive learning paradigm has a debiasing effect that can benefit recommender systems.
Specifically, our central contribution is establishing the theoretical connection between contrastive learning and inverse propensity weighting, where the latter is a well-studied technique for bias reduction~\citep{rosenbaum1983central,imbens2015causal,thompson2012sampling}.
Our theory complements the previous studies of contrastive learning~\citep{cpc}.

Based on the theoretical discovery, an easy-to-implement framework is developed to efficiently reduce the exposure bias of a large-scale system.
Our implementation maintains a fixed-size first-in first-out (FIFO) queue~\citep{moco} to accumulate positive samples and their representations from the previous batches, and use the content of the queue to serve as the negative samples to be discriminated from the positive samples of the next batch.
It guarantees that all items will be sampled sometime in an epoch to serve as the negative examples.
More importantly, it allows us to reuse the computed results from the previous batches, e.g., saving 90\% computation cost when the queue size is $10\times$ of the batch size.
As a result, we can afford to encode complex features of the negative samples even when the negative sample size, i.e., the queue size, is very large, where the rich features can improve the quality of the learned representations of the under-explored items.
The queue-based framework also offers simple yet efficient implementations for complex self-supervised tasks, e.g., discriminating whether two subsequences come from the same user's behaviors~\citep{jx-seq2seq}.
We further improve upon CLRec and propose Multi-CLRec, which employs a multi-queue design for more accurate user-intention aware bias reduction.

Our methods have been fully deployed into our live system as the default choice to serve billions of page views each day.
We observe that they are capable of recommending high-quality items that are largely neglected by most current ongoing systems, and consistently outperform the previous state-or-art baselines.
    \begin{figure*}[t]
    \centering
    \includegraphics[width=0.78\textwidth]{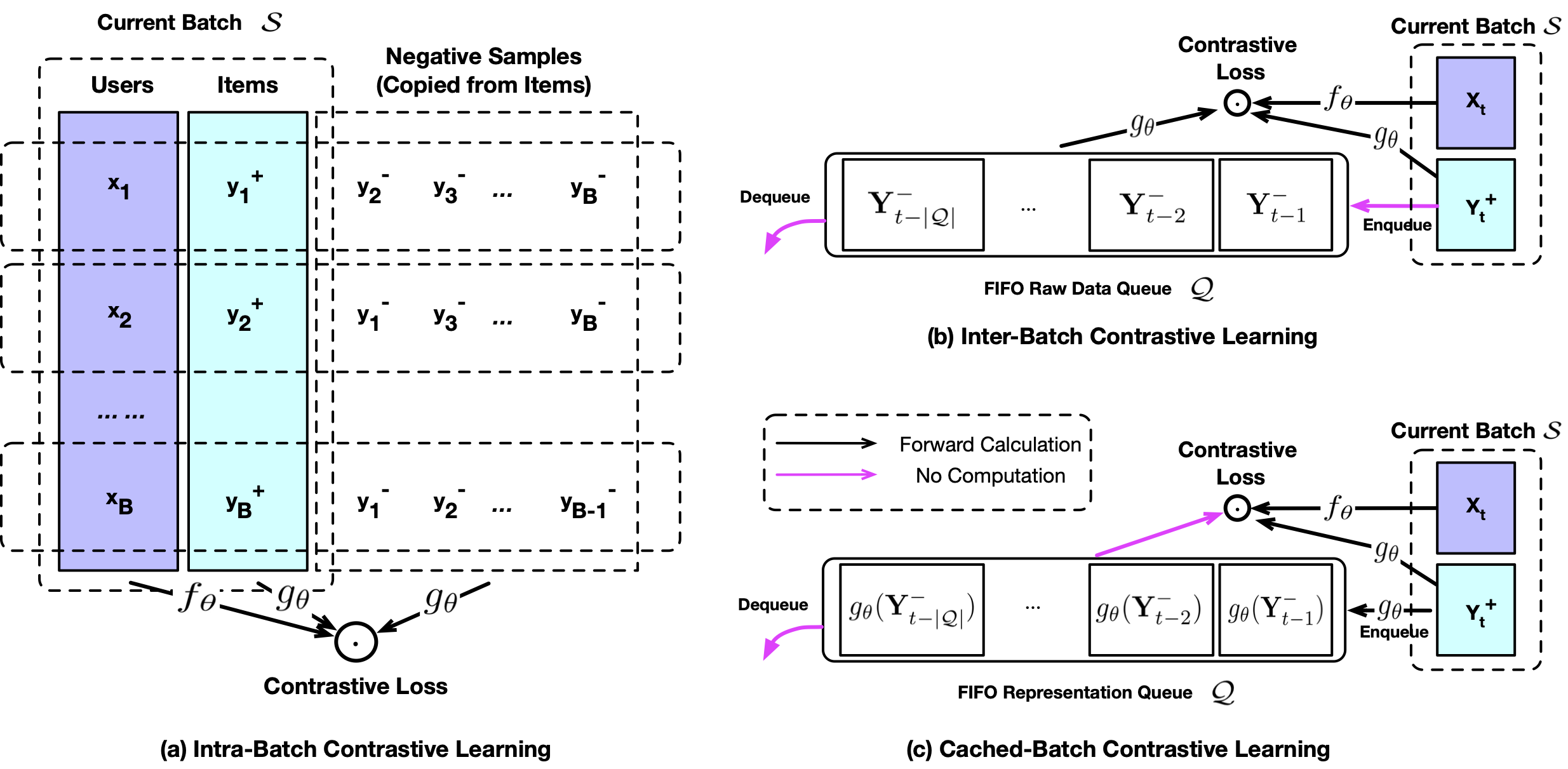}
    \caption{Three variants of the single-queue \methodname, whose implicit proposal distributions are $p_n(y\mid x)=p_{\mathrm{data}}(y)$.
    The superscripts $+,-$ mean positive and negative examples respectively.
    Variant~(a) uses the positive examples of other instances in the present batch as the negative examples.
    Variant~(b) creates a fixed-size FIFO queue to store the positive examples encountered in previously processed batches, and use the examples in the queue to serve as the negative examples for the present batch.
    Variant~(c) differs from variant~(b) in that the queue caches computed representations $\mathbf{g}_{\theta}(y)$ rather than raw features of $y$.}
    \label{fig:clrec}
\end{figure*}

\section{Our Theorectical Results on Contrastive Learning in DCG}
In this section, we will reveal the connection between contrastive learning and inverse propensity weighting (IPW).

\subsection{Problem Formulation}

\paragraph{Notations}
Given a dataset of user clicks $\mathcal{D}=\{(x_{u,t}, y_{u,t}): u=1,2,\ldots,N,\; t=1,2,\ldots,T_u\}$, where $x_{u,t}=\{y_{u,1:(t-1)}\}$
represents a user's clicks prior to the $t$-th click $y_{u,t}$, and $T_u$ denotes the number of clicks from the user $u$.
We will drop the sub-scripts occasionally and write $(x, y)$ in place of $(x_{u,t}, y_{u,t})$ for conciseness.
We use $\mathcal{X}$ to refer to the set of all possible click sequences, i.e.\ $x\in \mathcal{X}$.
Each $y\in\mathcal{Y}$ represents a clicked item, which includes various types of features associated with the item, while $\mathcal{Y}$ is the set of all possible items.
The features of $y$ could be in any form, e.g., the item's unique identifier number, embeddings or raw data of its image and text description.
The number of items $|\mathcal{Y}|$ can easily reach 100 million.

\paragraph{Deep Candidate Generation}
The deep candidate generation paradigm involves learning a user behavior encoder $\mathbf{f}_{\theta}(x)\in \mathbb{R}^d$ and an item encoder $\mathbf{g}_{\theta}(y)\in \mathbb{R}^d$.
The set of parameters used by each encoder is a subset of $\theta$, i.e.\ the set of all trainable parameters in the system.
It then takes $\{\mathbf{g}_{\theta}(y)\}_{y\in\mathcal{Y}}$ and builds a k-nearest-neighbor search service, e.g., Faiss~\citep{faiss}.
As a result, given an arbitrary user behavior sequence $x$ at serving time, we can instantly retrieve the top $k$ items relevant to the user by finding the top $k$ candidate $\mathbf{g}_{\theta}(y)$ similar to $\mathbf{f}_{\theta}(x)$.
Most implementations use inner product $\phi_\theta(x,y)=\langle \mathbf{f}_{\theta}(x), \mathbf{g}_{\theta}(y) \rangle$ or cosine similarity as the similarity score.
The typical learning procedure fits the data following the maximum likelihood estimation (MLE) principle:
\begin{equation}
    \begin{aligned}
        &\arg\min_\theta \frac{1}{|\mathcal{D}|}\sum_{(x,y)\in \mathcal{D}} -\log p_\theta(y \mid x),
        \quad \mathrm{where}\; \\
        & p_\theta(y\mid x)= \frac{\exp{\phi_\theta(x,y)}}{\sum_{y'\in \mathcal{Y}}{\exp{\phi_\theta(x,y')}}}.
    \end{aligned}
    \label{eq:full_softmax}
\end{equation}
The denominator of $p_\theta(y\mid x)$ sums over all possible items, which is infeasible in practice and thus requires approximation, e.g.,\ via sampling.
However, the observed clicks for training are from the previous version of the recommender system.
The training data thus suffer from exposure bias (i.e.\ missing not at random~\citep{schnabel2016debias}) and reflect the users' preference regarding the recommended items rather than all potential items.
High-quality items that have few clicks in the training data will likely remain under-recommended by a new algorithm trained via the MLE paradigm.

\subsection{Understanding Contrastive Learning from a Bias-Reduction Perspective}

We now introduce the family of contrastive losses that we are interested in, and reveal their connection with the inverse propensity weighting (IPW)
~\citep{rosenbaum1983central,imbens2015causal,thompson2012sampling} techniques for bias reduction.

\paragraph{Sampled Softmax}
The kind of contrastive loss we will investigate is similar to sampled softmax.
We thus recap sampled softmax here and will show that the minor difference is crucial.
There are many variants of sampeld softmax~\citep{sampledsoftmax_raw,sampledsoftmax}, among which the following variant is integrated by TensorFlow~\citep{abadi2016tensorflow} and commonly used:
\begin{equation}
    \begin{aligned}
        &  \arg\min_\theta \frac{1}{|\mathcal{D}|} \sum_{(x,y)\in \mathcal{D}} - \log  \\
        & \left\{\frac{\exp \left(\phi_\theta(x,y)-\log p_n(y| x)\right)}{
        \exp \left(\phi_\theta(x,y)-\log p_n(y| x)\right) + \sum_{i=1}^L \exp \left(\phi_\theta(x,y_i)-\log p_n(y_i| x)\right)
        }\right\},
    \end{aligned}
    \label{eq:sampled_softmax}
\end{equation}
where $\{y_i\}_{i=1}^L$ are $L$ negative samples drawn from a pre-defined proposal distribution $p_n(y\mid x)$.
Subtracting $\log p_n(y\mid x)$ is necessary for it to converge to the same solution as the exact loss in Eq.~(\ref{eq:full_softmax}).
Most implementations assume $p_n(y\mid x)=p_n(y)$ and set $p_n(y)$ somehow proportional to the popularity of the items to improve convergence.
In practice, we would draw thousands of negative samples to pair with each positive example.
Sampled softmax in general outperforms other approximations such as NCE~\citep{nce} and negative sampling~\citep{w2v} when the vocabulary is large~\citep{mccallum1998comparison,vae_cf_www,youtube,sampledsoftmax}.

\paragraph{Contrastive Loss}
We study the following type of contrastive loss~\citep{sohn2016npairmc,cpc} under a negative sampler $p_n(y\mid x)$:
\begin{equation}
    \arg\min_\theta \frac{1}{|\mathcal{D}|} \sum_{(x,y)\in \mathcal{D}} - \log
    \frac{\exp \left(\phi_\theta(x,y)\right)}{\exp \left(\phi_\theta(x,y)\right)+\sum_{i=1}^L \exp \left(\phi_\theta(x,y_i)\right)},
    \label{eq:clrec}
\end{equation}
where $\{y_i\}_{i=1}^L$ are again sampled from $p_n(y\mid x)$ for each $x$ to pair with the positive sample.
It no longer optimizes the MLE loss in Eq.~(\ref{eq:full_softmax}), because it misses $-\log p_n(y\mid x)$ and thus does not correct the bias introduced by sampling.
Many efforts on contrastive learning have been focusing on designing a well-performing proposal distribution $p_n(y\mid x)$~\citep{sohn2016npairmc,caselles2018word2vec}.
InfoNCE~\citep{cpc} demonstrates that this loss maximizes a lower bound of the mutual information between $x$ and $y$ if we set the proposal distribution
$p_n(y\mid x)$ to be the actual data distribution $p_{\mathrm{data}}(y)$, i.e.\ if we sample $y$ proportional to its frequency in the dataset.

\paragraph{Contrastive Learning and Exposure Bias Reduction}
The contrastive loss as shown above in Eq.~(\ref{eq:clrec}) has recently achieved remarkable success in various fields~\citep{moco,simclr,cpc}.
Nevertheless, it still remains a question why the loss is effective.
We will reveal that the contrastive loss is a sampling-based approximation of an inverse propensity weighted (IPW) loss.
The IPW loss~\footnote{
Our IPW loss is different from the previous works on debiased recommenders~\cite{schnabel2016debias,wang2016unbias_ipw}.
We focus on \emph{multinomial} propensities, i.e.\ whether an item is selected and recommended by a recommender out of all possible items.
The previous works consider \emph{bernoulli} propensities related with users' attention, i.e.\ whether a user notices a recommended item or not, and mostly deal with the position bias in ranking.
We give the derivation of this multinomial IPW loss in the Appendix. 
} has the form:
\begin{equation}
    \arg\min_\theta
    \frac{1}{|\mathcal{D}|} \sum_{(x,y)\in \mathcal{D}} -\frac{1}{q(y \mid x)} \cdot \log p_\theta(y \mid x),
    \label{eq:ipw_loss}
\end{equation}
where $q(y \mid x)$ should be the propensity score function, which represents the \emph{impression (or exposure) distribution}, i.e., the probability that item $y$ is recommended to user $x$ in the previous system where we collect the training data $\mathcal{D}$.
The idea of IPW is to model missing-not-at-random via the propensity scores in order to correct the exposure bias.
We can prove that the IPW loss is optimizing $p_\theta(y\mid x)$ to capture the oracle user preference even when there exists exposure bias.
A standard implementation of the IPW loss has two steps, where the first step is to use a separate model to serve as $q(y \mid x)$ and optimize it by fitting the exposure history according to the MLE principle, while the second step is to optimize $p_\theta(y \mid x)$ according to Eq.~(\ref{eq:ipw_loss}).
However, the two-stage pipeline of IPW, as well as the numerical instability and variance~\citep{schnabel2016debias} brought by $\frac{1}{q(y \mid x)}$, makes IPW less efficient for large-scale production systems.

Fortunately, we can prove that the contrastive loss Eq.~(\ref{eq:clrec}) is in principle optimizing the same objective as Eq.~(\ref{eq:ipw_loss}).
And in Subsection~\ref{sec:framework} we provide simple implementations that do not require two separate steps and can avoid the instability brought by the division of $q(y \mid x)$.
Our key theoretical result is as follows:
\begin{theorem}
    The optimal solutions of the contrastive loss (Eq.~\ref{eq:clrec}) and the IPW loss (Eq.~\ref{eq:ipw_loss}) both minimize the KL divergence from $p_\theta(y \mid x)$ to $r(y\mid x)=\frac{ p_\mathrm{data}{(y\mid x)} / q(y\mid x) }{ \sum_{y'\in \mathcal{Y}}  p_{\mathrm{data}}(y'\mid x) / q(y'\mid x) }$, if $p_n(y\mid x)$ is set to be $q(y\mid x)$.
    Here $p_{\mathrm{data}}(y\mid x)$ is the data distribution, i.e.\ what is the frequency of $y$ apprearing in $\mathcal{D}$ given context $x$.
    \label{theory:main}
\end{theorem}

\begin{proof}
    See the Appendix. 
\end{proof}

\begin{remark}
    The implication of Theorem~\ref{theory:main} is that the contrastive loss (Eq.~\ref{eq:clrec}) can approximately reduce the exposure bias if we set the proposal distribution $p_n(y\mid x)$ to be the propensity score, i.e.\ the probability that the old systems deliver item $y$ to user $x$ when we were collecting the training data $\mathcal{D}$.
\end{remark}

\section{Queue-based Contrastive Learning for Bias Reduction}\label{sec:framework}

\begin{figure*}[t]
\centering
\includegraphics[width=0.92\textwidth]{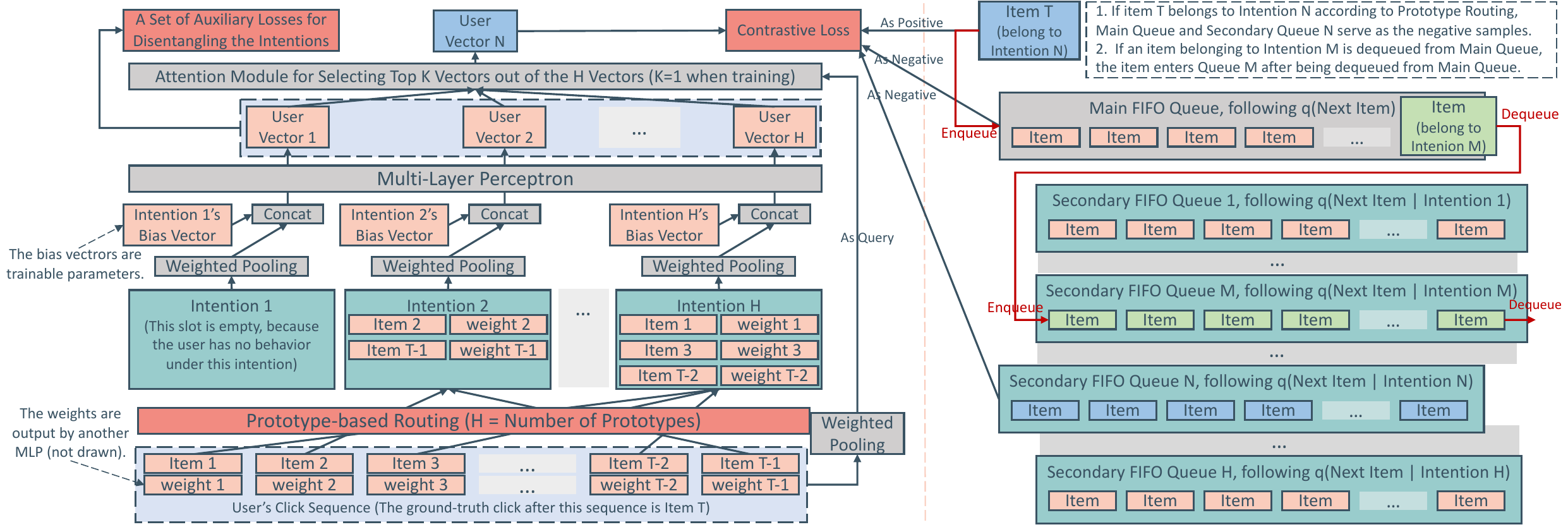}
\caption{Multi-CLRec, a multi-queue implementation of contrastive learning for bias reduction in DCG.}
\label{fig:multclr}
\end{figure*}

We propose two efficient implementations of contrastive losses for bias reduction in large-scale DCG:
(1) We will first present CLRec (see Figure~\ref{fig:clrec}), which draws negative samples from a queue that roughly follows the non-personalized propensity score $q(y)$, i.e., the probability of item $y$ being recommended by the old system under which we collect the training data, for reducing the Matthew effect.
(2) We then improve upon CLRec and propose Multi-CLRec (see Figure~\ref{fig:multclr}), which involves multiple disentangled queues corresponding to different user intents and performs bias reduction based on the more accurate propensity score $q(y\mid \textnormal{user $x$'s intent})$.

Our approaches, unlike the IPW methods that require introducing an extra model for estimating the propensity score $q(y\mid x)$, are free from the variance brought by the extra model and do not need to introduce a costly, extra stage for training the extra model.

\subsection{CLRec: Queue-based Contrastive Learning for Reducing the Matthew Effect}
\label{sec:single-queue-clrec}

The exact propensity score $q(y\mid x)$ is challenging to estimate, because industrial systems involve many complicated modules and the data are sparse.
CLRec thus use $q(y)$ in place of $q(y\mid x)$, i.e.\ assuming $q(y\mid x)\approx q(y)$, to ease estimation and avoid small propensities.
Secondly, $q(y)$ (i.e.\ the exposure probability that item $y$ is recommended to someone) has a high correlation with $p_{data}(y)$, i.e.\ the click probability that item $y$ is being recommended and clicked by someone, because the existing system will mainly recommend items that have a high click-through rate if the system is already highly optimized.
We thus further replace the impression distribution $q(y)$ with the click distribution $p_{data}(y)$, i.e.\ assuming $q(y)\approx p_{data}(y)$, to ease implementation
and save computation since impression data is larger than click data by an order of magnitude.

\paragraph{Drawbacks of Performing Explicit Sampling}
Our analysis above shows that, by simply drawing negative samples according to $p_{data}(y)$, the contrastive loss (Eq.~\ref{eq:clrec}) can alleviate the selection bias related with the items' impression counts in the history.
However, sampling will still incur non-negligible overheads, e.g.\ communication costs, in a distributed environment~\citep{negative_distribute}.
Sampling also cannot guarantee that every item will be sampled in an epoch.
We thus adopt a queue-based design~\citep{moco} that avoids explicitly performing sampling, as shown in Figure~\ref{fig:clrec}b and Figure~\ref{fig:clrec}c.

\paragraph{Designs of Queue-based Contrastive Learning}
We maintain a first-in first-out (FIFO) queue $\mathcal{Q}$, which has a fixed capacity and can store $|\mathcal{Q}|$ examples.
Given a new batch to process, we first enqueue the positive examples $y$ (or their representations $\mathbf{g}_\theta(y)$) encountered in the present batch into $\mathcal{Q}$.
We then use the examples stored in the queue to serve as $\{y\}\cup\{y_i\}_{i=1}^L$ (the positive example and $L$ negative examples) when computing the denominator of the contrastive loss (Eq.~\ref{eq:clrec}) for the present batch.
In a distributed setting, each worker maintains its own queue locally to avoid communication costs.
When the queue size $|\mathcal{Q}|$ is equal to the batch size, our implementation is then equivalent to sampling negative examples from the present batch~\citep{sohn2016npairmc,gru4rec,chen2017sampling} (see Figure~\ref{fig:clrec}a).
In general, we need thousands of negative samples to achieve satisfying performance.
We thus use a large queue size, but with a small batch size to save memory, e.g.\ batch size $=256$ and queue size $=2,560$.

\paragraph{Queue-based Caching and its Efficiency in Encoding Complex Features}
With the implementation that stores the encoded results $\mathbf{g}_\theta(y)$ in the queue (see Figure~\ref{fig:clrec}c), we can no longer back-propagate through the negative examples from the previous batches,
though we can still back-propagate through the negative examples from the present batch.
As a result, we find that the total training steps required for convergence mildly increase.
However, since each training step will take much less time to compute (and less communication costs in a distributed training setting), the total running time can still be greatly reduced if the features of the negative items are expensive to encode, e.g.\ if the features contain raw images, texts, or even structured data such as a knowledge graph.

\subsection{Multi-CLRec: Multiple Disentangled Queues for Intent-Aware Bias Reduction}
\label{sec-multclr}

There are tens of thousands of categories of items in our system, many of which are long-tail.
As a result, approximating $q(y\mid x)$ with $q(y \mid \textnormal{the category in which user $x$ is currently interested})$ can still incur non-negligible variance.
Multi-CLRec thus aims to cluster the categories into $H$ \emph{intention}s and simultaneously learn a de-biased model with negative samples that roughly follow the smoothed propensity scores $q(y\mid \textnormal{user $x$'s current intention is $h$})$, $h=1,\ldots, H$.
In particular, Multi-CLRec uses $H$ queues $\{\mathcal{Q}_h\}_{h=1}^H$ corresponding to the $H$ intentions.
We use $H=64$ in our system.

\paragraph{Disentangled Encoding with Prototype-based Routing}
Our model (see Figure~\ref{fig:multclr}) borrows the idea of intention disentanglement~\citep{disentangle_rec}, which has a built-in routing mechanism for clustering.
Let $x=\{y_{1:(T-1)}\}$ be the sequence of items clicked by user $x$, and the user's ground-truth next click to predict is $y_{T}$.
Let $\mathbf{c}_t\in \mathbb{R}^d$ be the embedding corresponding to item $y_t$'s category ID.
The embedding of item $y_t$ is generated by the item encoder $\mathbf{g}_\theta: \mathcal{Y} \to \mathbb{R}^d$ as follows:
\begin{equation}
\begin{aligned}
\mathbf{g}_\theta(y_{t})= \mathtt{MLP_1}([
& \textnormal{ embedding of item $y_t$'s unique ID};
\\ & \textnormal{ embedding of item $y_t$'s category ID, i.e., $\mathbf{c}_t$};
\\ & \textnormal{ embedding of item $y_t$'s seller ID};
\\ &\textnormal{ embeddings of $y_t$'s tags; \; \ldots } ]),
\end{aligned}
\end{equation}
where $\mathtt{MLP}_1(\cdot)$ represents a multilayer perceptron (MLP) whose input is the concatenation of item $y_t$'s features.
We then use another MLP to model the importance of each clicked item $\{y_t\}_{t=1}^{T-1}$:
\begin{equation}
\begin{aligned}
& p_{t} = \frac{\exp(p_{t}')}{\sum_{t'=1}^{T-1} \exp(p_{t'}')},
\;\; \textnormal{where} \\
p_{t}' = \mathtt{MLP_2}([
& \textnormal{ item embedding $\mathbf{g}_\theta(y_{t})$; category embedding $\mathbf{c}_t$};
\\ &\textnormal{ time gap between clicking item $y_t$ and item $y_T$};
\\ &\textnormal{ user's dwell time on item $y_t$; \;\; \ldots } ]) \; \textnormal{and $p_{t}'\in\mathbb{R}$}.
\end{aligned}
\end{equation}

Our model then uses a set of trainable parameters $\boldsymbol{\mu}_h\in \mathbb{R}^d, h=1,2,\ldots, H$, to represent $H$ intention prototypes~\citep{disentangle_rec}, based on which each item $y_t$ is being routed into intention $h$ with probability $p_{h\mid t}$:
\begin{equation}
\begin{aligned}
p_{h\mid t} = \frac{\exp(p_{h\mid t}')}{\sum_{h'=1}^H \exp(p_{h'\mid t}')},
\;\; \mathrm{where}\;
p_{h\mid t}' = \frac{
\langle \boldsymbol{\mu}_h, \mathbf{c}_t \rangle
}{
\rho \cdot \| \boldsymbol{\mu}_h \| \cdot \| \mathbf{c}_t \|
}.
\end{aligned}
\end{equation}
The temperature hyper-parameter $\rho$ is set to $\rho=0.07$ following previous work~\citep{moco}, and $\|\cdot\|$ stands for the $l2$-norm of a vector.
Based on the importance weights and the routing results, we obtain $H$ intention vectors $\{\mathbf{z}_h\}_{h=1}^H$ for the user:
\begin{equation}
\begin{aligned}
\mathbf{z}_h = \mathtt{MLP_3}(
[\mathbf{z}_h'; \boldsymbol{\beta}_h]),
\;\; 
\mathbf{z}_h' = \sum_{t=1}^{T-1} p_{h\mid t} \cdot p_{t} \cdot
\mathbf{g}_\theta(y_{t}),
\end{aligned}
\end{equation}
where $\boldsymbol{\beta}_h\in \mathbb{R}^d, h=1,2,\ldots, H$, are trainable parameters that represent the bias vectors of the intentions, which are shared by all users.
However, not all of the $H$ intention vectors would match the user's current interest, since it is possible to have $\sum_{t=1}^{T-1} p_{h\mid t} \cdot p_{t}\to 0$ for some $h$.
We thus use the following attention module to select the intention vector that is most likely to match the user's interest, and set the the user encoder $\mathbf{f}_\theta(\cdot)$'s output to be the selected vector $\mathbf{z}_{h^*}$:
\begin{equation}
\begin{aligned}
& \mathbf{f}_{\theta}(x)= \mathbf{z}_{h^*},
\;\; \textnormal{where}\;
h^* = {\underset{h\in \{1,2,\ldots,H\}}{\arg\max}} w_h, \;\;
w_h = \frac{\exp (\tilde{w}_h)}{\sum_{h'=1}^H \exp (\tilde{w}_{h'})},
\\
& \tilde{w}_h = \frac{
\langle \mathbf{m}, \mathbf{z}_h \rangle
}{
\rho \cdot \|\mathbf{m}\| \cdot \|\mathbf{z}_h\|
},\;\;
\mathbf{m} = \mathtt{MLP_4}\left(\sum_{t=1}^{T-1} p_t \cdot \mathbf{g}_\theta(y_{t})\right).
\end{aligned}
\end{equation}
We approximate the gradient of argmax via \emph{straight-through}~\citep{straightthrough}.
That is, we define $\ddot{w}_h = \mathrm{stop\_gradient}(I[h=h^*] - w_h) + w_h$, where $I[h=h^*] = 1$ if $h= h^*$ and $I[h=h^*] = 0$ otherwise.
We then use $\ddot{w}_{h}$ instead of $h^*$ when computing the loss (see Equation~\ref{eq:mult-loss}).

When serving online, we can select the top-$K$ ($1\le K \le H$) vectors from $\{\mathbf{z}_h\}_{h=1}^H$ with maximum values of $w_h$ for item retrieval, which is useful if a diverse recommendation list is preferred.

\paragraph{Auxiliary Losses for Enhancing Interpretability}
The encoder architecture alone cannot guarantee an interpretable routing mechanism in our large-scale setting.
We thus introduce three auxiliary losses to enhance the interpretability~\footnote{
Empirical results that demonstrate the enhanced interpretability are provided in the Appendix. 
}.
The first is to encourage $p_{h\mid t}$ to be polarized, i.e., $p_{h^+\mid T}\to 1$ and $p_{h'\mid T}\to 0$ for $h'\neq h^+$:
\begin{equation}
\begin{aligned}
\mathcal{L}_{\mathrm{aux},1} = -\log p_{h^+\mid T}, \quad \textnormal{where}\;
h^+ = {\underset{h\in \{1,2,\ldots,H\}}{\arg\max}} p_{h\mid T}.
\end{aligned}
\end{equation}

The second auxiliary loss is for pushing the correct intention vector $\mathbf{z}_{h^+}$ closer to the target item $y_T$ while making the incorrect intention vectors $\{\mathbf{z}_{h'}\}_{h':h' \neq h^+}$, far away from the target item:
\begin{equation}
\begin{aligned}
\mathcal{L}_{\mathrm{aux},2} = -\log \frac{\exp(s_{h^+,T})}{\sum_{h'=1}^H \exp(s_{h',T})},\quad
s_{h,T}=
\frac{
\langle \mathbf{z}_h, \mathbf{g}_{\theta}(y_T) \rangle
}{
\rho \cdot \| \mathbf{z}_h \| \cdot \| \mathbf{g}_{\theta}(y_T) \|
}.
\end{aligned}
\end{equation}

Our third auxiliary loss is for balancing the $H$ intentions, such that each is responsible for a roughly equal number of items:
\begin{equation}
\begin{aligned}
\mathcal{L}_{\mathrm{aux},3} = \sum_{h=1}^H \frac{1}{H} \cdot (\log \frac{1}{H} - \log \pi_h ),
\quad \pi_h = \mathbb{E}_{\mathcal{B}}[p_{h\mid t}],
\end{aligned}
\end{equation}
where $\mathbb{E}_{\mathcal{B}}$ means that the expectation is computed based on the items encountered in the present mini-batch $\mathcal{B}$.

\paragraph{Multi-Queue Contrastive Loss}
The main loss to optimize is:
\begin{equation}
\begin{aligned}
\mathcal{L}_{\mathrm{cl}}= - \log
\sum_{h=1}^{H} \ddot{w}_h\cdot
\frac{
\exp \left\{\cos(\mathbf{z}_h, \mathbf{g}_{\theta}(y_T))/\rho\right\}
}{
\sum_{h'=1}^{H}
\sum_{y'\in \mathcal{Q}_{0}\cup\mathcal{Q}_{h^+}} \exp \left\{\cos(\mathbf{z}_{h'},\mathbf{g}_{\theta}(y'))/\rho\right\}
},
\label{eq:mult-loss}
\end{aligned}
\end{equation}
where $
\cos(x,y)= {
\langle \mathbf{a}, \mathbf{b} \rangle
} / (
\| \mathbf{a} \| \cdot \| \mathbf{b} \|
)
$.
Here $\mathcal{Q}_0$ is the main queue, as described in Subsection~\ref{sec:single-queue-clrec}, and $\mathcal{Q}_{h^+}$ is the $h^+$-th secondary queue (one of the $H$ secondary queues) that stores negative samples associated with intention $h^+ = {\arg\max}_h \; p_{h\mid T}$.
We update the $1+H$ queues prior to computing the loss at each iteration as follows:
(1) The positive sample $y_T$ is enqueued into the main queue $\mathcal{Q}_0$ prior to computing the loss.
(2) An item $y^-$ just dequeued from the main queue $\mathcal{Q}_0$ is enqueued into the secondary queue $\mathcal{Q}_{h^-}$, where
$
h^- = {\arg\max}_h \; p_{h\mid y^-} = {\arg\max}_h \;
\langle \boldsymbol{\mu}_h, \mathbf{c}^- \rangle / (\rho \cdot \| \boldsymbol{\mu}_h \| \cdot \| \mathbf{c}^- \|)
$
and $\mathbf{c}^-$ is the category embedding of item $y^-$.
(3) Items dequeued from the secondary queues $\{\mathcal{Q}_h\}_{h=1}^H$ are discarded.

We minimize the total loss $\mathcal{L}_{\mathrm{cl}}+\lambda_1 \mathcal{L}_{\mathrm{aux},1}+\lambda_2 \mathcal{L}_{\mathrm{aux},2}+\lambda_3 \mathcal{L}_{\mathrm{aux},3}$.
We simply set $\lambda_1=\lambda_2=\lambda_3=1$, because we find that our method is robust with respective to a wide range of coefficients.

\begin{remark}
We are implicitly performing bias reduction based on a propensity score proportional to $q(y\mid \textnormal{user $x$'s intention is $h$}) + \alpha \cdot q(y)$ for a certain smoothing factor $\alpha > 0$, by using the negative samples from a secondary queue $\mathcal{Q}_h$ and the main queue $\mathcal{Q}_0$.
\end{remark}


\section{Experiment}

\subsection{Experiments in Production Envrionments}
The online experiments have lasted for at least four months, and our algorithm serves several scenarios with different traffic volumes.
Details can be found in the Appendix.
The total number of items is around 100 million.
We use the queue-based implementation without caching in this subsection, and we will explore settings where encoding is expensive and requires caching in Subsection~\ref{sec:u2u}.

\begin{table}[t]

        \centering
        \caption{Aggregate diversity~\citep{diversity_aggregate}, i.e.\ the number of distinct items recommended to a randomly sampled subset of users.}
        \begin{tabular}{lr}
            \toprule
            &  {\bf Aggregated Diversity}  \\ \midrule
            sampled-softmax & 10,780,111 \\
            CLRec & 21,905,318 \\
            \bottomrule
        \end{tabular}
        \label{exp-aggdiversity}
\end{table}
\begin{figure}
\centering
        \includegraphics[width=0.43\textwidth]{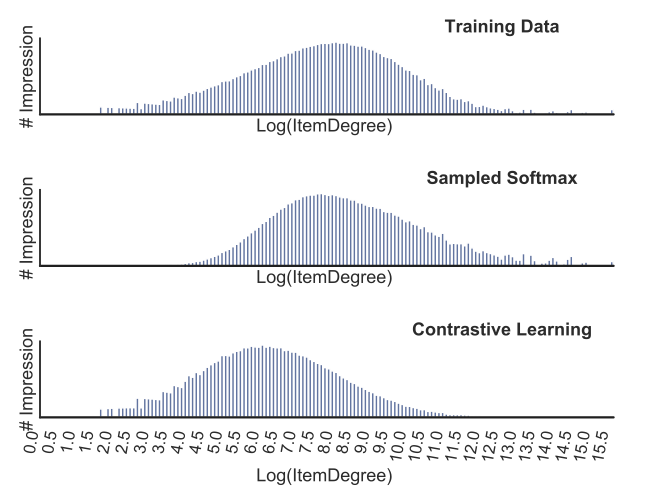}
        \captionof{figure}{
            The total number of impressions of the items in a specific degree bucket vs.\ the logarithm of the corresponding degree.
            The rightmost bar is not the highest because the number of the extremely popular items is small, even though each item in the bucket has a very high degree.}
        \label{fig:fairness}
\end{figure}

\begin{table}[t]
        \centering
        \caption{CLRec vs.\ the sampling-based alternatives.
        We conducted these proof-of-concept live experiments in a small-traffic scenario, due to the costs of online experiments.
        The negative sampling baseline has been outdated and removed from our live system before this work starts.
        }
        \begin{tabular}{@{}l@{}r@{}r@{}}
            \toprule
            {\bf Method} & {\bf HR@50 } & {\bf CTR(online)} \\ \midrule
            negative sampling & 7.1\% & outdated \\
            shared negative sampling & 6.4\% & -   \\
            sampled-softmax & 17.6\% & 3.32\% \\
            CLRec & { 17.8}\% & {3.85}\% \\
            \bottomrule
        \end{tabular}
        \label{exp-sampling}
\end{table}

\subsubsection{The Debiasing Effect in Large-Scale Production Environments}
\label{sec:exp-debias}
To examine the debiasing effects of CLRec, we first conduct offline experiments and compare CLRec with sampled softmax.
We report the aggregated diversity~\citep{diversity_aggregate} in Table~\ref{exp-aggdiversity}, and the distributions of the recommended items resulted from the different losses in Figure~\ref{fig:fairness}.

Table \ref{exp-aggdiversity} shows that \methodname has an over $2\times$ improvement on aggregated diversity.
We see from Figure \ref{fig:fairness} that, sampled softmax tends to faithfully fit the distribution of the training data.
CLRec, however, learns a relatively different distribution, which shifts towards the under-explored items and alleviates the Matthew effect.

Bias reduction not only leads to a fairer system, but also contributes to a significant improvement regarding the online performance.
In Table~\ref{exp-sampling}, we compare \methodname with other sampling alternatives using the same implementation of the encoders.
Details of the alternative methods are in the Appendix.
We observe that negative sampling~\cite{w2v}, including its variant~\cite{pinterest} that makes the instances in a batch share the same large set of negative samples, does not perform well in our settings. 
CLRec's improvement over sampled-softmax~\cite{sampledsoftmax,sampledsoftmax_raw} w.r.t.\ the offline metric HitRate@50 is negligible.
However, CLRec achieves significant improvement regarding the click-through rate (CTR) online, which indicates that there exists discrepancy between the offline metric and the online performance.

In Table~\ref{exp-private-multclr}, we compare Multi-CLRec against CLRec.
We deliberately evaluate the methods in the setting where the training and the test data have different user distributions.
Multi-CLRec is more robust compared to CLRec and achieves higher performance.

\begin{table}[t]
    \centering
    \caption{Main live experiment results conducted in one of the largest scenarios on our platform.
    CLRec consistently outperforms the baseline for months and has been fully deployed since Feb 2020.
    A method with a large popularity index tends to recommend popular items while a method with a small index is fairer to the under-exposured items.
    }
    \label{exp-mainlive}
    \begin{tabular}{@{}llcl@{}}
        \toprule
        {\bf Method} & {\bf CTR} & {\bf Average Dwell Time} & {\bf Popularity Index} \\ \midrule
        MIND & 5.87\%  & - & 0.658 \\ 
        CLRec & 6.30\%  & +11.9\% & 0.224 \\ 
        \bottomrule
    \end{tabular}
\end{table}

\begin{table}[t]
    \centering
    \caption{
    The user distribution of our platform usually changes significantly when the weekend arrives.
    When the user distribution changes significantly, Multi-CLRec has an advantage over single-queue CLRec in terms of the offline evaluation metric HitRate@50.
    Multi-CLRec replaced CLRec and became the new default in mid-2020, bringing a 3\% relative improvement in terms of total clicks on our platform while maintaining a comparable click-through rate.}
    \label{exp-private-multclr}
    \begin{tabular}{@{}ccc@{}}
        \toprule
        & \multicolumn{2}{c}{{\bf Train on Weekdays' Data}} \\
        \cmidrule(l){2-3}
        {\bf Method} & {\bf Test on Weekdays} & {\bf Test on Weekends}  \\ \midrule
        CLRec        & 17.18\%  & 17.16\% \\
        Multi-CLRec  & 17.25\%  & 17.68\% \\
        \bottomrule
    \end{tabular}
\end{table}

\subsubsection{At Least Four Months' Large-Scale A/B testing}
CLRec has been fully depolyed into several heavy-traffic scenarios since Feb 2020, after the initial proof-of-concept ablation studies shown in Table~\ref{exp-sampling}.
Table~\ref{exp-mainlive} shows our main online results conducted in these heavy-traffic scenarios, with billions of page views each day.
During the at least four months' A/B testing,
\methodname has been consistently outperforming MIND~\citep{mind}, the previous state-or-art baseline, in terms of the fairness metrics such as aggregated diversity and average popularity index, as well as the user engagement metrics such as click-through rate and average dwell time. 
Compared with MIND~\citep{mind}, which is the previous state-of-art DCG baseline deployed in the system,
CLRec tends to recommend items with a lower popularity index while being more attractive to the users.
This proves CLRec's ability of identifying high-quality items that are rarely explored by the previous systems.
We further achieve a $+2$\% relative improvement in the total click number on our platform after ensembling CLRec and MIND\eat{under the same number constraints of the recalled candidates}, compared to using MIND alone.

In mid-2020, we upgrade CLRec to Multi-CLRec, and Multi-CLRec brings a 3\% relative improvement in terms of total clicks on our platform while maintaining a comparable click-through rate.


\subsection{Computational Advantages of CLRec}
\subsubsection{Training Efficiency}
Table~\ref{exp-time} compares CLRec and sampled-softmax in terms of training speed and the network traffic required in a distributed setting.
CLRec's queue-based implementation is much more efficient than the methods that perform explicit sampling, since CLRec reuses the result computed for a positive sample shortly later when the sample is serving as a negative sample.
The version of sampled-softmax that encodes features for the negative items is from \cite{aligraph}.
This proof-of-concept experiment only uses a few features of the items, and we can deduce that the improvement regarding efficiency will be much more substantial with more complex features.
Table~\ref{exp-feature} shows that encoding features for the negative samples is beneficial, which justifies the efforts spent on efficiency.

\begin{table}[t]
    \centering
    \caption{
    Training speed in terms of the number of the positive examples processed per second, and
    the average network traffic in a distributed environment.
    }
    \label{exp-time}
    \begin{tabular}{@{}lcc@{}}
        \toprule
        {\bf Method} & {\bf Samples/Sec}  & {\bf NetTraffic (MB/s)} \\ \midrule
        sampled-softmax w/o feat. & $\approx 280k$ & $\approx 700$\\
        sampled-softmax with feat. & $\approx 130k$ & $\approx 1,100$\\
        CLRec w/o feat. & $\approx 330k$ & $\approx 500$\\
        CLRec with feat. & $\approx 280k$ & $\approx 500$\\
        \bottomrule
    \end{tabular}
\end{table}

\begin{table}[t]
        \centering
        \caption{
            The benefits of encoding features for the negative samples.
            Most baselines that employ sampled-softmax do not encode rich features for the negative samples (though they still use features when encoding the users' click sequences), because the number of negative samples is large and brings high costs if the features are complex.
            Fortunately, CLRec's cached implementation greatly reduces the costs, as demonstrated in Table~\ref{exp-time}.
        }
        \label{exp-feature}
        \begin{tabular}{@{}l@{}c@{}}
            \toprule
            {\bf Method} & {\bf HR@50} \\ \midrule
            CLRec + negatives w/o features & 17.4\%   \\
            CLRec + negatives with features & 19.4\%  \\
            \bottomrule
        \end{tabular}
\end{table}

\begin{table}[t]
        \centering
        \caption{
            Task u2i is the regular task where $x$ is a sequence of clicks and $y$ is the next click to be predicted.
            Task u2u~\citep{jx-seq2seq} adds an auxiliary loss where $x$ and $y$ are both sequences from the same user (before and after a sampled timestamp), which is co-trained with task u2i.
            HR1@50 and Recall5@50 represent the HitRate and Recall truncated at 50 when predicting the next one and five clicks, respectively.
        }
        \label{exp-seq2seq}
        \begin{tabular}{@{}l@{}cc@{}}
            \toprule
            \bf{Task \& Implementation} &  {\bf HR1@50} & {\bf Recall5@50} \\ \midrule
            CLRec-u2i & 17.9\% & 12.1\% \\
            CLRec-u2u, cached & 18.3\% & 12.7\% \\
            CLRec-u2u, cached + MoCo & 18.2\% & 12.6\% \\
            \bottomrule
        \end{tabular}
\end{table}

\begin{table}[t]
    \centering
    \caption{Results on public benchmarks preprocessed by SASRec\citep{sasrec} and BERT4Rec~\cite{bert4rec}.
    We follow BERT4Rec's approach when constructing the test set, which penalizes false positive predictions on popular items.
    For fair comparision, the IPW implementation and the CLRec implementation in this table use the same Transformer~\cite{self_attention} encoder as SASRec but with a IPW loss and a contrastive loss, respectively.
    }
    \label{exp-public-data}
    \begin{tabular}{ccccc}
        \toprule
        {\bf Method} & {\bf Metric} & {\bf ML-1M} & {\bf Beauty} & {\bf Steam} \\ \midrule
        SASRec & HR@1 & 0.2351 & 0.0906 & 0.0885  \\
        BERT4Rec & & 0.2863 & 0.0953 & 0.0957 \\
        IPW      & & 0.2934 & 0.1129 & 0.1381 \\
        \methodname & & \textbf{0.3013} & \textbf{0.1147} & \textbf{0.1424}  \\
        \midrule
        SASRec & HR@5 & 0.5434 & 0.1934 & 0.2559  \\
        BERT4Rec & & 0.5876 & 0.2207 & 0.2710 \\
        IPW      & & 0.5985 & 0.2404 & 0.3539 \\
        \methodname & & \textbf{0.6045} & \textbf{0.2552} & \textbf{0.3640}  \\
        \midrule
        SASRec & HR@10 & 0.6629 & 0.2653 & 0.3783  \\
        BERT4Rec & & 0.6970 & 0.3025 & 0.4013 \\
        IPW      & & 0.7156 & 0.3215 & 0.4924 \\
        \methodname & & \textbf{0.7194} & \textbf{0.3423} & \textbf{0.5019} \\
        \bottomrule
    \end{tabular}
\end{table}

\begin{table}[t]
    \centering
    \caption{Results on public benchmarks preprocessed by HPMN~\citep{ren2019lifelong}.
    We follow HPMN~\citep{ren2019lifelong} and use AUC as the metric.
    The clicks in the training and the test set of each benchmark happen before and after a certain timestamp, respectively.
    There hence exists a distribution drift.
    The users also have diverse interests, i.e., the probability of a user's two consecutive clicks belonging to the same category is low.
    }
    \label{exp-public-data-hpmn}
    \begin{tabular}{@{}ccc@{}}
        \toprule
        {\bf Method} & {\bf Amazon} & {\bf UserBehavior}\\ \midrule
        DNN    & 0.7546       & 0.7460       \\
        SVD++~\citep{koren2008factorization}  & 0.7155       & 0.8371       \\
        GRU4Rec~\citep{gru4rec}     & 0.7760       & 0.8471       \\
        Caser~\citep{tang2018personalized}   & 0.7582       & 0.8745       \\
        DIEN~\citep{dien}    & 0.7770       & 0.8934       \\
        RUM~\citep{chen2018sequential}     & 0.7464       & 0.8370       \\
        LSTM~\citep{hochreiter1997long}    & 0.7765       & 0.8681       \\
        SHAN~\citep{ying2018sequential}    & 0.7763       & 0.8828       \\
        HPMN~\citep{ren2019lifelong}    & 0.7809       & 0.9240       \\
        CLRec   & 0.7785       & 0.9278       \\
        IPW Loss {\small + Multi-CLRec's Encoder }  & 0.7888  & 0.9319                  \\
        Multi-CLRec & \textbf{0.7993    } & \textbf{0.9353    } \\
        \bottomrule
    \end{tabular}
\end{table}

\subsubsection{Complex Pretext Task that Requires the Cached Implementation}
\label{sec:u2u}
We now demonstrate that CLRec with a queue that caches the computed results can enable more complex pretext tasks that may improve the quality of the learned representations.
We consider an auxiliary task where $x$ and $y$ are both sequences from the same user (before and after a sampled timestamp).
The goal is to identify the correct sequence $y$ that belongs to the same user that produces $x$.
This auxiliary task is expensive to implement with sampled-softmax, since the negative samples are sequences and are thus expensive to encode.
Fortunately, cached CLRec can implement it efficiently.
Table~\ref{exp-seq2seq} demonstrates that the auxiliary task can improve an algorithm's ability to make long-term prediction.

MoCo~\cite{moco} proposes momentum update for stabilizing the training loss.
We however observe no additional gain with MoCo. 

\subsection{Reproducible Experiments on Public Data}

In addition to the experiments on large-scale data, we also conduct quantitative experiments on several public benchmarks. 
The quantitative results are listed in Table~\ref{exp-public-data} and Table~\ref{exp-public-data-hpmn}, respectively.

\begin{table}[t]
    \centering
    \caption{Aggregate diversity on public datasets, where each algorithm retrieves ten items for each test user.}
    \label{exp-aggregate-public}
    \begin{tabular}{rrrrr}
        \toprule
        & & \multicolumn{3}{c}{{\bf Aggregate Diversity}} \\
        \cmidrule(l){3-5}
        {\bf Dataset} & {\bf \#Users} & {\bf \small Ground-Truth} & {\bf \small Sample-Softmax} & {\bf \small CLRec} \\ \midrule
        ML-1M         & 6,040         & 1,883                    & 2,348                & 3,020                      \\
        Beauty        & 40,226        & 19,820                   & 24,340               & 44,408                     \\
        Steam         & 281,428       & 10,006                   & 5,154                & 10,111                     \\ \bottomrule
    \end{tabular}
\end{table}
\begin{figure}[t]
    \centering
    \subfigure[ML-1M.]{\includegraphics[width=0.15\textwidth]{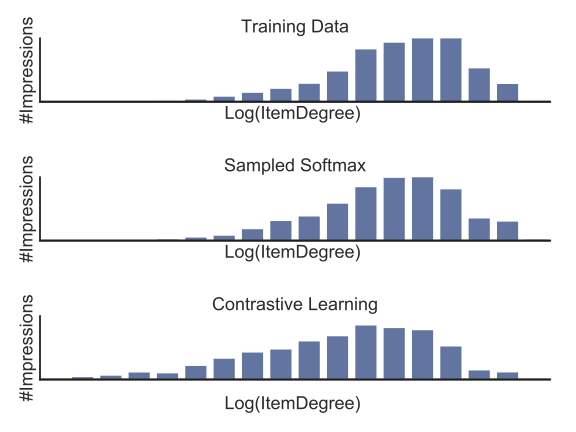}}
    \subfigure[Beauty.]{\includegraphics[width=0.15\textwidth]{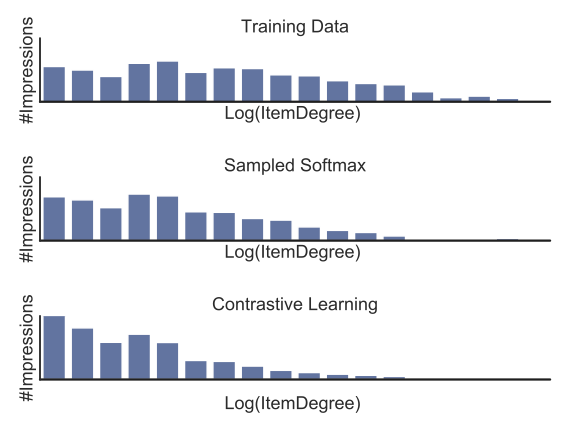}}
    \subfigure[Steam.]{\includegraphics[width=0.15\textwidth]{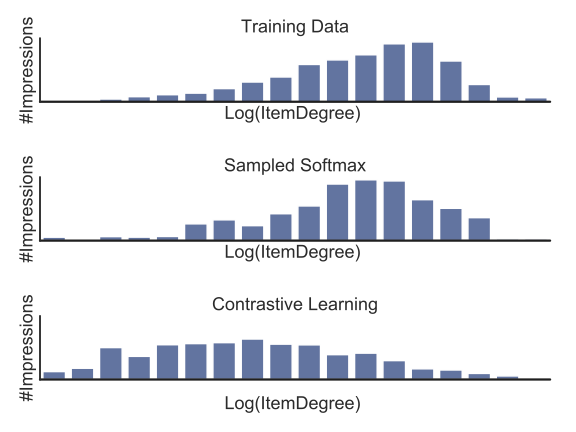}}
    \caption{The total number of impressions of the items a specific degree bucket vs.\ the logarithm of the corresponding degree.
    Sample softmax tends to faithfully fit the training data's distribution, while CLRec reduces the Matthew effect.}
    \label{exp-histogram-public}
\end{figure}

We also conduct the same analyses as in Subsection~\ref{sec:exp-debias} to demonstrate the debiasing effects of CLRec on the public benchmarks.
We sample a sequence from each user in the training set as the input to the recommenders.
Each recommender then retrieves ten items for each sequence.
The results are in Table~\ref{exp-aggregate-public} and Figure~\ref{exp-histogram-public}.

Table~\ref{exp-aggregate-public} reports aggregate diversity of the items retrieved by the algorithms, as well as the aggregate diversity of the ground-truth next clicks.
CLRec retrieves a more diverse set of items than sampled softmax, i.e., much more items will have a chance to be recommended after we reduce the bias in the training data.

Figure~\ref{exp-histogram-public} visualizes the distributions of the items retrieved by each algorithm, as well as the training data's distribution.
Sample softmax tends to faithfully fit the training data's distribution, while CLRec tends to explore the previously under-explored items.


\section{Related Work}

\paragraph{Deep Candidate Generation}
Deep candidate generation methods are widely deployed in industrial systems, e.g., YouTube~\citep{youtube,joglekar2019neural,topkoffpolicy}, Taobao~\citep{app,mind,disentangle_rec,tdm}, and Pinterest~\citep{pinterest}.
The existing methods explicitly sample negative examples from a pre-defined proposal distribution~\citep{w2v,sampledsoftmax,sampledsoftmax_raw}.
The proposal distribution not only affects convergence, but also has a significant impact on the performance~\citep{caselles2018word2vec}.
Empirically the number of the negative samples need to be large, e.g.\ a few thousand ones for pairing with a positive example.
Consequently, it is computationally expensive to incorporate rich features for the negative samples.
The existing systems hence usually choose to not encode features for the negative examples except for simple features such as item IDs~\citep{youtube}, even though rich features for the negative samples are demonstrated to be beneficial~\citep{bai2019bundle}.
CLRec achieves great efficiency when encoding rich features for the negative samples by caching the computed results.

\paragraph{Bias Reduction and Fairness in Recommender Systems}
Recommendation algorithms that directly fits the training data will suffer from selection bias due to the missing-not-at-random phenomenon~\citep{schnabel2016debias}, where the previous recommendation algorithms affect the training data collected.
The topic of reducing the bias in training and evaluating recommender systems has been explored before~\citep{steck2013evaluation,ai2018unbiased,wang2016unbias_ipw,schnabel2016debias,topkoffpolicy,yang2018unbiased,causerec,bundletreat,hy-stablegraph}.
However, these existing works mostly focus on small-scale offline settings, and rely on techniques less efficient for large-scale DCG.
For example, most of them involve an extra stage to train a propensity score estimator.
Dividing the propensity scores can also lead to high variance when small propensities exist~\citep{schnabel2016debias}.
Correcting the bias helps improve P-fairness~\citep{burke2017multisided}, i.e.\ fairness towards the under-recommended products~\citep{fairness_kdd}.


\paragraph{Contrastive Learning}
Contrastive learning, which aims to learn high-quality representations via self-supervised pretext tasks, recently achieves remarkable success in various domains, e.g., speech processing~\citep{cpc}, computer vision~\citep{simclr,deepinfomax,moco}, graph data~\cite{graphinfomax}, and compositional environments~\citep{worldmodel}.
The contrastive loss we investigate in this paper is a generalization of the InfoNCE loss~\citep{cpc}.
InfoNCE is previously understood as a bound of the mutual information between two variables~\citep{cpc}.
Our work provides a new perspective on the effectiveness of the contrastive loss, by illustrating its connection with inverse propensity weighting~\citep{rosenbaum1983central,imbens2015causal,thompson2012sampling}.

\section{Conclusion}
We established in theory the connection between contrastive learning and IPW, based on which we proposed CLRec and Multi-CLRec for efficient and effective bias reduction in large-scale DCG.


    \bibliographystyle{ACM-Reference-Format}
    \balance
    \bibliography{reference_short}

    \appendix
    \section{Experimental Settings}


\paragraph{Data and Hyper-parameters}
We use the data collected from the last four days on our platform as the training data and use the click data recorded in the following day as the test set.
We limit the minimum length of a valid sequence to be five in the training set, because requiring the model to predict the next item of an unusually short sequence can easily introduce great noise which harms the recommendation performance.
We use seven categorical features of the items, including item ID, coarse-grained and fine-grained category IDs, seller ID, seller category, brand ID, and gender preference, for our live experiments.
We use the queue-based implementation without the caching mechanism on the regular tasks where the negative items consist of the seven categorical features, while using the queue-based implementation that caches the processed results when solving the complex pretext tasks due to the high costs of encoding the complex negative examples.
The number of negative samples is 2,560 while the batch size is 256.
Multi-CLRec use $H=16$ on public data and $H=64$ on our large-scale data.
We train the model for one epoch in total on our large-scale data.

\paragraph{Training Environment}
The training data contain four billion sequences of user behaviors.
We train the model in a distributed TensorFlow cluster that consists of 140 workers and 10 parameter servers.
Each worker is equipped with a GPU, which is either GTX 1080 Ti or Tesla P100.
Each parameter server is allocated 32GB memory while each worker has 16G memory.
Training CLRec and Multi-CLRec on our internal four-day data for one epoch costs less than eight hours for task u2i without caching and less than twelve hours for task u2u with caching.

\paragraph{Online Serving}
We pre-compute offline the item representations via the item encoder.
We then store and index the item representations using an online vector-based kNN service for fast retrieval of top relevant items at serving time.
At serving time, our user encoder will infer a user vector $\mathbf{f}_\theta(x)$ as the request made by user $x$ arrives and pass the user vector to the kNN service to find a few hundred relevant items $S$.
The latency for dealing with the query is around 10 ms.
The set of relevant items $S$ is then send to be scored by a separate deep ranking model that costs much more computational power and is responsible for constructing the final recommendation list by selecting the top relevant items from $S$.

\paragraph{Evaluation Metrics}
We report the performance in terms of the following offline and online metrics:
\begin{itemize}
    \item HitRate@50 (HR@50) is an offline metric.
    We sample a subset of users and compute the metric based on the users' click sequences to measure an algorithm's offline performance.
    We pick the latest one click of a user as the label $y$ to predict, and use the user' sequence prior to $y$ to serve as the input $x$.
    We run the algorithm under evaluation to retrieve top 50 items for each user via kNN searching.
    If the ground-truth next one click $y$ is in the retrieved 50 items, the algorithm receives a score $1$ for this user, or score $0$ otherwise.
    The HitRate@50 of the algorithm is then the average score received by the algorithm.
    \item The click-trough rate (CTR) of an algorithm is the percentage of recommendations made by the algorithm that are finally clicked by the users.
    \item The average dwell time is the total time spent by the users on reading the details of the items clicked by them, divided by the total number of clicks on our platform.
    \item The aggregate diversity~\citep{diversity_aggregate}, measured on a sampled subset of users for testing, is the number of distinct items recommended to the subset of users.
    \item The popularity index of the items recommended by a candidate generation algorithm $A$ is defined as $\frac{\mathbb{E}_A[p_{\mathrm{data}}(y)]}{\max_{A'} \mathbb{E}_{A'}[p_{\mathrm{data}}(y)]}$.
    Here $p_{\mathrm{data}}(y)$ is the number of times our system recommends item $y$ to the users in the past few weeks before we deploy CLRec into the system, and
    $\mathbb{E}_A[p_{\mathrm{data}}(y)] = \sum_{y\in \mathcal{Y}} p_{\mathrm{data}}(y)\cdot p(\mathrm{algorithm}\; A\; \mathrm{recommends\; item}\; y)$ measures how much an algorithm prefers recommending the items that are already popular.
    The algorithm $A'$ that has the largest value of $\mathbb{E}_{A'}[p_{\mathrm{data}}(y)]$ in our system is a simple candidate generation method that does not learn vector representations, which is used as a fallback to handle the cases when the deep candidate generation methods, i.e.\ MIND and CLRec, fail to return the results within a time limit.
\end{itemize}

\section{Implementation Details}

\subsection{Single-Queue CLRec's Encoders}

The encoders of the multi-queue implementation Multi-CLRec is described in Subsection~\ref{sec-multclr}.
However, the single-queue CLRec uses a different set of user and item encoders, which is due to the fact that our large-scale experiments related with the single-queue CLRec were conducted before we invented Multi-CLRec.
We thus describe in this subsection the single-queue CLRec's encoders.

\subsubsection{Item Encoder}
In the single-queue CLRec,
an item's embedding is the sum of several parts, i.e.\ a base item embedding corresponding to its ID umber, the concatenation of the item features' embeddings, a time bucket embedding, as well as a backward positional embedding.
The time bucket embedding and the positional embedding are only added when the item embedding is to be used by the user encoder $\mathbf{f}_\theta(x)$ for encoding a past click made by user $x$.
The output of the item encoder $\mathbf{g}_{\theta}(y)$ for item $y$ is the sum of only the base item embedding and the item features' embeddings, excluding the time and positional embeddings.
Notably, Multi-CLRec also use the time and positional embeddings described here for estimating each click's importance weight.

\paragraph{Time Bucket Embedding}
To encode time intervals between consecutive behaviors, we use the embeddings of bucketized time intervals, which are used by previous work~\citep{atrank}.
To be specific, given the current timestamp $t$ when the user query arrives and the timestamp $t'$ when a click was recorded, the time interval $t-t'$ is discretized and put into a bucket that corresponds to a specific range.
Each bucket has its own vector representation, i.e.\ bucket embedding.

\paragraph{Backward Positional Embedding}
We also find that the positional embedding has extra positive impacts even when the time bucket embedding is already in use.
There are two ways to add positional embeddings.
The forward way is to view the latest click as if it is at the anchor position, i.e.\ position zero, while the earliest click of the user would then be at a varying position depending on the user's sequence length.
The backward way is, on the other hand, to view the earliest click as if it is at position zero, and the latest click would then be at a varying position when the sequence length varies.
We find that the backward version of the positional embedding brings extra improvement even after the bucket embedding is used, while the forward positional embedding does not.
We thus adopt the backward positional embedding.
The time bucket embedding has already highlighted the recent behaviors, and the backward positional embedding may be behaving like the backward direction in a bi-directional LSTM for capturing long-range dependencies.



\subsubsection{User Encoder}

For single-queue CLRec, we employ a simplified multi-head attention (MultiAtt) module for capturing a user's diverse interests hidden behind the user's click sequence $x=\{y_t\}_{t=1}^{T-1}$.
We then perform weighted head aggregation (WHA) of the multiple heads' outputs, i.e., head embeddings.
That is,
$
\mathbf{f}_{\theta}(x)=\mathtt{WHA}\left(
\mathtt{MultiAtt}\left(\mathbf{X}\right)
\right)
$,
where $\mathbf{X}\in \mathbb{R}^{(T-1)\times d}$ are the $T-1$ clicks' embeddings, whose specification has been described earlier.

\paragraph{Multi-Head Attention}
Multi-head attention~\citep{google_attention} achieves great success in various NLP tasks and is adopted by sequential recommendation models~\citep{sasrec,atrank}.
We simplify the multi-head attention~\citep{self_attention} for efficiency's sake and implement the following module:
\[
    \mathtt{MultiAtt(X)}=\mathtt{SoftMax}(\mathtt{MLP}(\mathbf{X})^\top, -1)\;\mathbf{X},
\]
where $\mathbf{X}\in \mathbb{R}^{(T-1)\times d}$ are the $T-1$ clicks' embeddings.
$\mathtt{MLP}$ is a two-layer feed forward neural network whose output is of shape $\mathbb{R}^{(T-1)\times H}$, representing the $H$ sets of attention scores for the $H$ attention heads, respectively.
The softmax is performed along the last axis of its input, hence the -1.
The output of the above module is a set of $H$ $d$-dimensional embeddings, namely $H$ head embeddings.

\paragraph{Weighted Heads Aggregation (WHA)}
After the simplified multi-head attention module, there are $H$ head embeddings $\{\mathbf{z}_h\}_{h=1}^H$.
We use a WHA module to aggregate the $H$ head embeddings into one single user embedding.
WHA first produces a global feature vector $\mathbf{m}\in\mathbb{R}^d$, based on which another weighted attention is performed to produce the final user embedding $\mathbf{f}_{\theta}(x)\in\mathbb{R}^d$ for user $x$:
\begin{equation*}
    \label{eq:WHA}
    \mathbf{m}=\frac{\sum_{h=1}^{H}\mathbf{z}_h}{H},\;
    \alpha_h=\frac{\exp(
    \mathbf{z}_{h}^\top \mathbf{A} \mathbf{m}
    )}{\sum_{h'=1}^H{\exp(
    \mathbf{z}_{h'}^\top \mathbf{A} \mathbf{m}
    )}},\;
    \mathbf{f}_{\theta}(x)=\sum_{h=1}^H \alpha_h \mathbf{z}_{h},
\end{equation*}
where $\mathbf{A}\in\mathbb{R}^{d\times d}$ is a learnable parameter.
We find WHA to be more effective than mean-pooling or concatenation.

\subsection{Normalization and Initialization}

WHA and contrastive learning both seem to be sensitive to the similarity metrics used.
We thus $l2$-normalize the head embeddings before sending them into WHA and normalize the output of WHA as well, which empirically improves training convergence and recommendation diversity.
We use the cosine similarity with temperature $\rho > 0$ for contrastive learning, i.e.
$
\phi_\theta(x, y)=\frac{
\langle \mathbf{f}_{\theta}(x), \mathbf{g}_{\theta}(y) \rangle
}{\rho \|\mathbf{f}_{\theta}(x)\| \|\mathbf{g}_{\theta}(y)\|}
$,
where $\rho=0.07$ following previous work~\citep{moco}.

Some efforts~\citep{wang2018cosface,wojke2018deep,disentangle_rec} have been spent on analyzing the cosine similarity related with $l2$-normalization as well as the positive effects brought by $l2$-normalization.
According to \cite{wojke2018deep} which assumes a normal distribution, cosine softmax will force the embeddings to form more compact clusters, which makes kNN search more easy.

The embeddings should use normal initialization~\citep{glorot2010} instead of uniform initialization if $l2$-normalization is in use, so that the initial embeddings are uniformly spread over the hyper-sphere, which prevents the loss from oscillating at the beginning of training.


\subsection{Loss Functions and Sampling Strategies}
\label{sec-appendix-loss}

\paragraph{Negative sampling~\citep{w2v}}
We sample $L$ examples to pair with each positive click from a proposal distribution $q(y)$.
The proposal distribution $q(y)$ is proportional to the item's degree, i.e., $\mathrm{degree}(y)^{0.75}$ as recommended in \cite{w2v}.
We use a distributed version of the sampling strategy based on the alias method~\citep{negative_distribute}, which is provided by AliGraph~\citep{aligraph}.
We tune $L$ from $8$ to $2,560$ and report the best.

\paragraph{Shared negative sampling~\citep{pinterest}}
This variant aims to increase the efficiency of the original implementation of negative sampling.
It makes the positive examples in the present batch share the same set of $L$ negative samples, rather than sampling a different set of $L$ negative samples for each positive instance.
Sampled softmax, as well as CLRec, similarly let the positive examples in a batch share the same set of negative samples.
However, this implementation of negative sampling with sharing still uses the binary cross-entropy loss used by the original negative sampling implementation.

\paragraph{Sampled softmax~\citep{sampledsoftmax}}
We use the sampled softmax implemented in Tensorflow~\citep{abadi2016tensorflow}, which subtracts the correction term to make the sampled loss approximately optimize the same objective as the full loss.
We set the number of negative examples used by sample softmax per batch to be the same as the queue size of CLRec.

\paragraph{CLRec \& Multi-CLRec}
The batch size is 256.
The queue size of CLRec is 2,560.
The queue size of Multi-CLRec is 1,280, leading to 2,560 negative items in total as it draws samples from both a main queue and an additional secondary queue.

\subsection{Complex Pretext Tasks}
\label{sec-appendix-u2u}

In task u2u~\citep{jx-seq2seq}, $x$ and $y$ are both sequences from the same user, before and after a sampled timestamp, respectively.
Intuitively, pre-training a recommender to solve task u2u may improve a recommender's ability to make long-term prediction.
We co-train the same encoders to solve both task u2u and the original u2i task where $x$ and $y$ are a sequence and the sequence's next click.
Task u2i is still required, since we ultimately need the recommender to recommmend items rather than sequences at serving time.
The batch size is 256.
We use one separate queue of size 2,560 for each task.
We limit the sequence length of $y$ in task u2u to be less than ten.
We also investigate whether Momentum Contrast (MoCo)~\citep{moco} may improve performance, and vary the momentum parameter in the range of $\{0.9, 0.99, 0.999, 0.9999\}$ and report the best result.

    \section{Additional Empirical Results}

In this section, we provide additional empirical results that demonstrate the interpretability of Multi-CLRec and justify some designs of our encoders.

\subsection{Results about Multi-CLRec's Encoders}

\begin{figure}[t]
    \centering
    \includegraphics[width=0.30\textwidth]{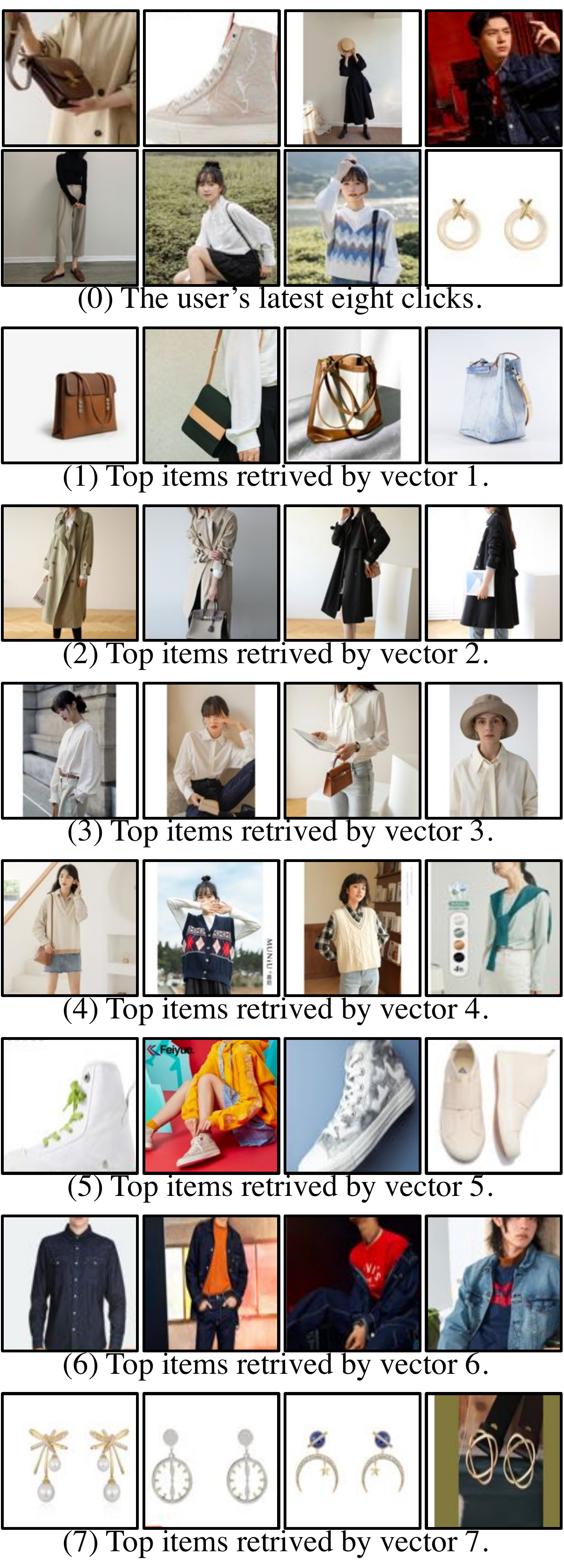}
    \caption{Multi-CLRec's user encoder can select and output top-$K$ vectors at serving time.
    We show here the top-$K$ vectors (for $K=7$) do represent the user's $K$ different intentions, owing to the auxiliary losses for intention disentanglement.
    }
    \label{fig:case-multivec}
\end{figure}

In Figure~\ref{fig:case-multivec}, we demonstrate that the top-$K$ vectors selected by Multi-CLRec's user encoder do represent a user's $K$ different intentions, which is mostly due to the auxiliary losses for intention disentanglement.
We note that the case shown here is \emph{not} cherry-picked, but a randomly sampled case.

\subsection{Results about CLRec's Encoders}

In this subsection, we provide some interesting empirical findings that motivates the design of our encoders when developing the single-vector CLRec, even though the findings are not strictly related to contrastive learning or bias reduction.
Each of the following experiments should be taken as the analysis for each component individually, which are conducted during different time periods.
Considering the costs of deployment into a live system, we deploy a model online only if it has at least similar performance with the previous baselines, and mark a method's online performance as ``-'' in the table if it is not deployed.

Table~\ref{exp-posemb} then demonstrate the usefulness of combining time bucket embedding and backward positional embedding.

We then illustrate in Table~\ref{exp-headaggr}that the encoders can benefit from the combination of weighted head aggregation (WHA) and the cosine similarity, where the results show that cosine similarity is indeed needed for reaching a higher HitRate.

\begin{figure}[h]
    \centering
    \subfigure[User's behavior sequence.]{\includegraphics[width=0.42\textwidth]{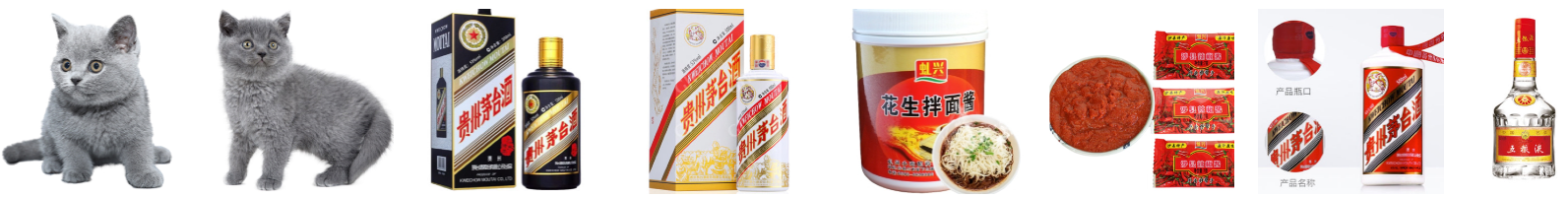} \label{fig:case-trigger} }
    \subfigure[Items retrieved by the concatenation method.]{\includegraphics[width=0.42\textwidth]{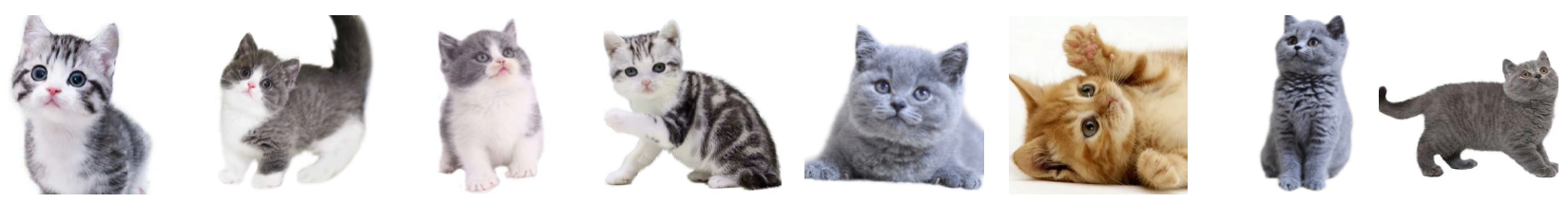}   \label{fig:case-ss}}
    \subfigure[Items retrieved by the WHA + cosine method.]{\includegraphics[width=0.42\textwidth]{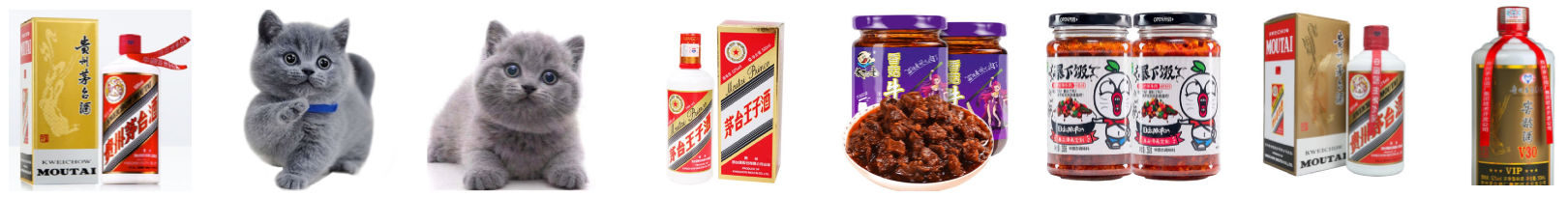} \label{fig:case-cl} }
    \vspace{-0.2cm}
    \caption{It is possible for a single-vector user encoder to generate a diverse recommendation list, as long as the model is properly designed, e.g., if the model combines the weighted head aggregation (WHA) strategy with the cosine similarity.
    Nevertheless, the single-vector CLRec is not as controllable and reliable as the multi-vector Multi-CLRec when it comes to preserving the multiple intentions in a balanced manner.
    }
    \label{fig:case}
    \vspace{-0.3cm}
\end{figure}

We also find that combining WHA and cosine can endow a single-vector user encoder with the ability to generate a diversified recommendation list that reflects the different intents behind a user's behavior sequence (see Figure~\ref{fig:case}).
Without WHA, we observe that the retrieved item set can be easily dominated by one single intent, showing limited diversity.

The previous methods, especially those use dot product instead of cosine similarity, easily lead to a much skewer interest distribution of the top-k retrieved items.
These previous methods tend to under-explore items from the less popular interests.
Many have also noticed this issue and proposed their solutions, such as using multiple interest vectors~\citep{disentangle_rec,mind} or multiple interest sub-models~\citep{tdm}.
Our result (Figure~\ref{fig:case}), however, indicates that a single-vector model can sometimes be sufficient for generating a diverse recommendation list.

\begin{table}[h]
    \centering
    \caption{Effects of the forward positional embedding, backward positional embedding, and time bucket embedding~(TBE).
    We conducted these ablation studies about the encoders in a small-traffic scenario.
    }
    \vspace{-0.15cm}
    \label{exp-posemb}
    \begin{tabular}{@{}lcc@{}}
        \toprule
        {\bf Method}         & {\bf HR@50 (Offline)} & {\bf CTR (Online)} \\ \midrule
        FwdPosEmb            & 18.5\%                & -                  \\
        BackPosEmb           & 18.2\%                & 2.98\%             \\
        TimeBucketEmb        & 18.9\%                & -                  \\
        FwdPositionEmb + TBE & 16.4\%                & -                  \\
        BackPosEmb + TBE     & 19.1\%                & 3.25\%             \\
        \bottomrule
    \end{tabular}
    \vspace{-0.15cm}
\end{table}

\begin{table}[h]
    \centering
    \caption{Performance of the different head aggregation strategies used by the single-vector CLRec's user encoder.}
    \vspace{-0.15cm}
    \label{exp-headaggr}
    \begin{tabular}{@{}lcc@{}}
        \toprule
        {\bf Method}               & {\bf HR@50 (Offline)}                  & {\bf CTR (Online)}  \\ \midrule
        MeanPooling                & 15.1\%                                 & 3.10\%              \\
        MeanPooling + Cosine       & 18.1\%                                 & -                   \\
        MeanPooling + MLP          & 10.5\%                                 & -                   \\
        MeanPooling + MLP + Cosine & 17.7\%                                 & -                   \\
        Concatenation + MLP        & 17.8\%                                 & 3.15\%              \\
        WHA + Cosine               & 19.7\%                                 & 3.21\%              \\
        \bottomrule
    \end{tabular}
    \vspace{-0.45cm}
\end{table}

    \section{Theorectical Analyses and Proofs}

In this section, we provide the derivation of the IPW loss for a multinomial policy, along with the proof of our theorem.

\subsection{Derivation of Inverse Propensity Weighting for Multinomial Policies}
\label{sec:app-ipw}

In this subsection, we derive the IPW loss for DCG, where we represent the recommendation policy as a multinomial distribution.
We assume that the recommender system is a sequential recommender system where a user will receive only one recommendation $y$ when the user's state becomes $x$.
However, we note that it is easy to verify that our conclusions still hold when user state $x$ receives $K$ recommendations, where $K \ge 1$.
We will also focus on a single user state $x$ for conciseness.

\subsubsection{Training with Unbiased Data}

Ideally, we should collect the training data that do not suffer from selection bias.
This is equivalent to collect training data under a random recommendation policy $\pi_\mathrm{uni}$ that always recommends a random item in a uniform way, i.e., the random policy should not be aware of the value of $x$ and $y$.

Let the oracle user preference be $p(\mathrm{click}=1|x,y)$, which is a bernoulli distribution and represents how likely user $x$ will click item $y$ if we recommend item $y$ to the user.
Assuming that user $x$ clicks one item recommended by the random policy $\pi_\mathrm{uni}$, the click data collected under $\pi_\mathrm{uni}$ will be:
\begin{equation}
    p_{\pi_\mathrm{uni}}(y|x) = \frac{p(\mathrm{click}=1\mid x,y)}{\sum_{y'}p(\mathrm{click}=1\mid x,y')},
    \label{eq:opt_pi}
\end{equation}
which is the oracle click data distribution for unbiased training, since the uniform random policy $\pi_\mathrm{uni}$ does not introduce any selection bias.
Note that $p_{\pi_\mathrm{uni}}(y|x)$ is a multinomial distribution rather than a multivariate bernoulli distribution.

To learn an ubiased recommendation policy, we need to optimize $p_\theta(y|x)$ on the click data collected under the uniform random policy.
In other words, the unbiased loss function should be:
\begin{equation}
    R(\theta|x)=-\sum_y p_{\pi_\mathrm{uni}}(y|x)\log p_\theta(y|x).
    \label{eq:goal}
\end{equation}

\subsubsection{Training with Biased Data: Inverse Propensity Weighting}

However, random exploration is expensive in a real-world recommender system.
In practice, the click data $\mathcal{D}_\pi$ for training a new policy is collected under a biased old recommendation policy $\pi$.
Let $q_\pi(y\mid x)$ be the probability that the old policy recommends item $y$ to a user in state $x$.
Note that $q_\pi(y\mid x)$ is a multinomial distribution, i.e.\ $\sum_{y'}q_\pi(y'\mid x)=1$.

The generating process of the biased dataset $\mathcal{D}_\pi$ is as follows.
Policy $\pi$ makes a recommendatin, i.e.\ draws a one-hot impression vector $O$ from the multinomial distribution $q_{\pi}(y\mid x)$, whose $y$th element $O_y=1$ if the recommendation is item $y$, and $O_y=0$ otherwise.
On the other hand, user $x$ is associated with a multi-hot vector $C$ representing the user's preference regarding all the items, whose $y$th element $C_y$ is drawn from the bernoulli distribution $p(\mathrm{click}=1\mid x,y)$, $y=1,2,3,...,|\mathcal{Y}|$.
Here $C_y=1$ if the user will click item $y$ after being recommeded item $y$, $C_y=0$ otherwise.

A na\"ive estimator that does not take selection bias into account typically learns a new biased policy $p_\theta(y|x)$ by minimizing the following biased loss:
\begin{equation}
    \hat{R}{_\mathrm{naive}}(\theta|\mathcal{D}_\pi)
    =-\sum_{y}O_y C_y\log{p_\theta(y|x)},
\end{equation}
Since $O$ and $C$ are independent, the expectation of the above loss is
\begin{align*}
    \mathbb{E}_{\mathcal{D}_\pi}\left[\hat{R}_\mathrm{naive}(\theta|\mathcal{D}_\pi)\right]
    & =-\sum_{y}\mathbb{E}[{O_y}]\mathbb{E}[{C_y}]\log{p_\theta(y\mid x)} \\
    & =-\sum_{y} q_\pi(y\mid x) p(\mathrm{click}=1\mid x,y)  \log{p_\theta(y\mid x)}.
\end{align*}
The na\"ive estimator is thus not optimizing the ideal loss as described in Eq.~(\ref{eq:goal}).

Now let us analyze the IPW estimator, which minimizes the following loss:
\begin{equation}
    \hat{R}_\mathrm{IPW}(\theta|\mathcal{D}_\pi) =-\sum_{y}\frac{1}{q(y\mid x)}{O_y}{C_y}\log{p_\theta(y\mid x)}.
    \label{eq:estimator}
\end{equation}
Note that $p_{\pi_\mathrm{uni}}(y|x) \propto p(\mathrm{click}=1\mid x,y)$ according to Eq.~(\ref{eq:opt_pi}).
We thus have:
\begin{align*}
    \mathbb{E}_{\mathcal{D}_\pi}\left[\hat{R}_\mathrm{IPW}(\theta|\mathcal{D}_\pi)\right]
    & = -\sum_{y}\frac{q_\pi(y\mid x)}{q(y\mid x)}  p(\mathrm{click}=1\mid x, y) \log {p_\theta(y\mid x)} \\
    & \propto -\sum_{y}\frac{q_\pi(y\mid x)}{q(y\mid x)}  p_{\pi_\mathrm{uni}}(y\mid x) \log {p_\theta(y\mid x)}.
    \label{eq:exp_ipw_new}
\end{align*}
We can see that, if we set $q(y|x)$ to $q_\pi(y|x)$, the IPW estimator is then an unbiased estimation that minimizes the oracle unbiased loss Eq.~(\ref{eq:goal}).

\subsubsection{Bernoulli Propensities vs.\ Multinomial Propensities}

We note that our fomulation is different from the existing literature on debiasing a recommender~\citep{steck2013evaluation,ai2018unbiased,wang2016unbias_ipw,schnabel2016debias,topkoffpolicy,yang2018unbiased}.
Most of the existing methods assume that the recommendations $O$ are drawn from a multivariate bernoulli policy, i.e., $O_y$ follows an independent bernoulli distribution $q_\pi(\mathrm{recommend}=1\mid x,y)$~\citep{wang2016unbias_ipw,schnabel2016debias}.
Here $O_y\in\{0,1\}$ denotes whether item $y$ is recommended by the policy.
We take a different approach and instead assume that $O$ is drawn from a multinomial policy $q_\pi (y\mid x)$ rather than a multivariate bernoulli one.

In other words, the difference lies in whether we should design a candidate generation policy as a multinomial model or as a multivariate bernoulli model.
We choose to design the policy as a multinomial one, due to the following reasons:
\begin{itemize}
    \item \emph{The multinomial formulation brings superior performance}.
    There are many empirical results that report better performance with a multinomial candidate generation method than a multivariate bernoulli one, especially with a large set of items for recommendation~\citep{youtube,mind,sampledsoftmax,vae_cf_www}.
    Our empirical results in Table~\ref{exp-sampling}, where the negative sampling baselines follow the multivariate bernoulli formulation, also verify this finding.
    Similarly, the natural language processing (NLP) community also report better results with a multinomial classifier, especially when the number of classes increases~\citep{mccallum1998comparison}.
    \item
    \emph{The multinomial formulation is more consistent with the real-world production environments.}
    The multivariate bernoulli formulation implicitly assumes that the number of recommendations requested by a user can be modeled as part of the recommendation policy $\pi$, e.g., the expectation of the number of recommendations is $\sum_y q_\pi(\mathrm{recommend}=1\mid x,y)$ with the multivariate bernoulli formulation.
    However, this assumption does not hold in many recommender systems, where the number of recommendations received by a user is decided by the user rather than the system.
    For example, the user can request more recommendations by scrolling down the page or stop receiving any new recommendation by leaving the page.
    On the contrary, the multinomial policy $q_\pi (y\mid x)$ does not attempt to model the number of recommendations requested by a user.
    Rather, the multinomial policy $q_\pi (y\mid x)$ only models which one item it should recommend if the user explicitly requests the system to make one recommendation.
\end{itemize}


\subsection{Proof of Theorem 1}

\begin{theorem}
    The optimal solutions of the contrastive loss (Eq.~\ref{eq:clrec}) and the IPW loss (Eq.~\ref{eq:ipw_loss}) both minimize the KL divergence from $p_\theta(y \mid x)$ to $r(y\mid x)=\frac{ p_\mathrm{data}{(y\mid x)} / q(y\mid x) }{ \sum_{y'\in \mathcal{Y}}  p_{\mathrm{data}}(y'\mid x) / q(y'\mid x) }$, if $p_n(y\mid x)$ is set to be $q(y\mid x)$.
    Here $p_{\mathrm{data}}(y\mid x)$ is the data distribution, i.e.\ what is the frequency of $y$ apprearing in $\mathcal{D}$ given context $x$.
\end{theorem}

\begin{proof}
    We now give a proof sketch on the theorem.
    We will focus on one training instance, i.e.\ one sequence $x\in\mathcal{X}$.
    The IPW loss (Eq.~\ref{eq:ipw_loss}) for training sample $x$ is
    \begin{equation}
        \begin{aligned}
            &-\sum_{y:(x,y)\in\mathcal{D}}\frac{
            \log p_\theta(y \mid x)
            }{q(y \mid x)}
            \propto -\sum_{y\in \mathcal{Y}} \frac{p_{\mathrm{data}}(y\mid x)}{q(y \mid x)} \log p_\theta(y \mid x) \\
            &\propto -\sum_{y\in \mathcal{Y}} r(y \mid x) \log p_\theta(y \mid x)
            =D_{\mathrm{KL}}
            (r \| p_\theta) + \mathrm{const. w.r.t.}\; \theta.
        \end{aligned}
    \end{equation}
    The IPW loss is thus minimizing the Kullback–Leibler (KL) divergence from $p_\theta(y \mid x)$ to $r(y\mid x)$.

    Let us now focus on the contrastive loss for the training sample $(x,y)$.
    Let $C=\{y\}\cup \{y_i\}_{i=1}^L$, where $y$ is the positive example and $\{y_i\}_{i=1}^L$ are the $L$ negative samples drawn from $q(y\mid x)$.
    Note that $C$ is a multi-set where we allow the same item to appear multiple times.
    The contrastive loss (Eq.~\ref{eq:clrec}) for $x$ equals to
    \begin{equation}
        \begin{aligned}
            &- \sum_{y:(x,y)\in \mathcal{D}}
            \sum_{C} q(C\mid x,y)
            \log
            \frac{\exp \left(\phi_\theta(x,y)\right)}{\sum_{y'\in C} \exp \left(\phi_\theta(x,y')\right)}
            \\
            \propto &
            - \sum_{y\in \mathcal{Y}}
            \sum_{C} q(C\mid x,y)
            p_{data}(y\mid x) \log
            \frac{\exp \left(\phi_\theta(x,y)\right)}{\sum_{y'\in C} \exp \left(\phi_\theta(x,y')\right)},
        \end{aligned}
    \end{equation}
    where $q(C\mid x,y)=\prod_{i=1}^L q(y_i\mid x)$ if $y\in C$ or $q(C\mid x,y)=0$ if $y\notin C$, since by definition $C$ must include $y$ if we know that the positive example is $y$.

    Let $q(C\mid x)= \prod_{y'\in C} q(y'\mid x)$.
    We then have $ q(C\mid x,y)= \frac{q(C\mid x)}{q(y\mid x)}$ if $C$ includes $y$.
    As a result, we can see that the contrastive loss for training sample $x$ is proportional to
    \begin{equation}
        \begin{aligned}
            &-
            \sum_{y\in \mathcal{Y}}
            \sum_{C:y\in C}
            \frac{q(C\mid x)}{q(y\mid x)} p_{data}(y\mid x)
            \log
            \frac{\exp \left(\phi_\theta(x,y)\right)}{\sum_{y'\in C} \exp \left(\phi_\theta(x,y')\right)}
            \\
            = & \mathbb{E}_{q(C\mid x)}\left[
            -
            \sum_{y\in C}
            \frac{p_{data}(y\mid x)}{q(y\mid x)}
            \log
            \frac{\exp \left(\phi_\theta(x,y)\right)}{\sum_{y'\in C} \exp \left(\phi_\theta(x,y')\right)}
            \right]
            \\
            = &
            \mathbb{E}_{q(C\mid x)}\left[
            D_{\mathrm{KL}}
            (r^C \| p_\theta^C)
            \right]
            + \mathrm{const. w.r.t.}\; \theta.
        \end{aligned}
    \end{equation}
    Here we use $r^C$ and $p_\theta^C$ to refer to the probability distributions
    $r^C(y| x)=\frac{ p_{\mathrm{data}}(y\mid x) / q(y\mid x) }{ \sum_{y'\in C}  p_{\mathrm{data}}(y'\mid x) / q(y'\mid x) }$
    and
    $p_\theta^C(y| x)=\frac{\exp \left(\phi_\theta(x,y)\right)}{\sum_{y'\in C} \exp \left(\phi_\theta(x,y')\right)}$, whose supports are $C\subset \mathcal{Y}$.
    Since we are minimizing the KL divergence under all possible $C\subset\mathcal{Y}$, the global optima will be the ones that make $p_\theta(y \mid x)$ equal to $r(y\mid x)$ for all $y\in \mathcal{Y}$ if $\phi_\theta(x,y)$ is expressive enough to fit the target distribution arbitrarily close.
    Note that $\phi_\theta(x,y)$ is indeed expressive enough since we implement it as a neural network, due to the universal approximation theorem~\citep{cybenko89,hornik91}.
    The two losses hence have the same global optima.
\end{proof}


\end{document}